\documentclass[reprint,superscriptaddress,amsmath,amssymb,prl]{revtex4-1}

\usepackage{graphicx}
\usepackage{subfigure}
\usepackage{dcolumn}
\usepackage{bm}
\usepackage{xcolor}%
\usepackage{amsmath,amsthm}
\usepackage{verbatim}
\usepackage{mathtools}
\usepackage{float}

\def\nin{\noindent}
\def\beq{\begin{equation}}
\def\eeq{\end{equation}}

\begin{document}

\title{Sink-rise dynamics of horizontally oscillating active matter \\ in granular media: Theory}


\author{Liu Ping} 
\affiliation{National University of Defense Technology, Changsha 410073, China}
\affiliation{Cavendish Laboratory, JJ Thomson Avenue, Cambridge CB3 0HE, UK}
\author{Xianwen Ran} 
\affiliation{National University of Defense Technology, Changsha 410073, China}
\author{Raphael Blumenfeld} \email{rbb11@cam.ac.uk}
\affiliation{Cavendish Laboratory, JJ Thomson Avenue, Cambridge CB3 0HE, UK}

\date{\today}

\begin{abstract}

An intermediate step to modelling behaviour of active matter is understanding interactions of active objects (AOs) with inanimate matter, which often lead to a range of rich behaviour. We present a range of simulations of the interaction of a self-energised AO with three-dimensional granular matter and develop a first-principles theoretical model to describe the observed phenomena.
The AO oscillates horizontally, which causes it to either rise against gravity or sink, depending on the oscillation amplitude and frequency. We identify two competing mechanisms that drive the vertical motion. When the AO moves below a critical speed, $v_c$, it generates a jammed stagnant zone ahead of it, which effects an upward force and leads to the rise. Above $v_c$ and certain kinetic energy, the medium around the AO fluidises and the AO sinks into the layer supporting it. The duration of the rising and sinking  phases depend non-trivially on the AO's amplitude and frequency leading to an intricate nonlinear dynamics.
We derive the equation of motion for the time-dependent depth from first-principles and show that its solutions agree well with a wide range of computer simulations, which we perform within the range of parameters allowed by the finiteness of the simulated system.

\end{abstract}

\maketitle

Active objects (AOs) possess internal energy sources that can be used to generate movement by applying forces on surrounding media. An AO may be a part of a collection, e.g. schools of fish, flocks of birds, or bacterial colonies, interacting with other AOs \cite{CollectiveAOs}, or it may interact with inanimate media, such as lizards, scorpions, and snakes burrowing and `swimming' in sand \cite{SandAOs}. While interacting many-AO populations have attracted much attention, these are complex and system-specific. In contrast, AO-inanimate matter interactions, while also leading to rich behaviours, are more amenable to theoretical modelling.
Here, we formulate a first-principles theoretical model for the intriguing dynamics of animals in granular media.
The model comprises an AO submerged in a granular medium and executing a  self-energised horizontal periodic oscillation.
This issue is relevant to specific applications in robotics \cite{Robotics} and animal locomotion in sand \cite{locomotion}, as well as to general understanding of interactions of AOs with passive particulate media.

The dynamics of intruders in granular media require understanding the combined effects of gravity, drag and lift forces. While drag on objects moving in granular media has been studied in several contexts \cite{Aletal99,Drag,Kangetal18,Fengetal19}, effects of forces acting normally to the direction of motion have been less explored \cite{LiftForce}. Recent simulations of two-dimensional systems showed that horizontally oscillating AOs may either rise against gravity or sink, depending on the oscillation amplitude and frequency \cite{Huetal16}. A cavity model was proposed to explain the mechanisms driving this phenomenon, but it fell short of leading to an equation of motion (EoM) and predict the AO's trajectories.
Here, we first extend the work to three dimensions (3D), using simulations at a wide range of amplitudes and frequencies and varying the AO's density, initial depth, size and friction coefficient. We find that the cavity model does not capture correctly the dynamics. 
We propose that the rich dynamics result from a competition between lift forces, driven by formation of a stagnant zone (SZ) ahead of the AO, and gravity, as the layer supporting it fluidises partially at high velocities. Based on our observations, we derive an EoM for the AO's depth and solve it. The good agreement between the theoretical predictions and the simulation results supports strongly our theory.\\

\nin \textbf{Simulation procedure}:
\begin{figure}[H]
\includegraphics[width=0.48\textwidth]{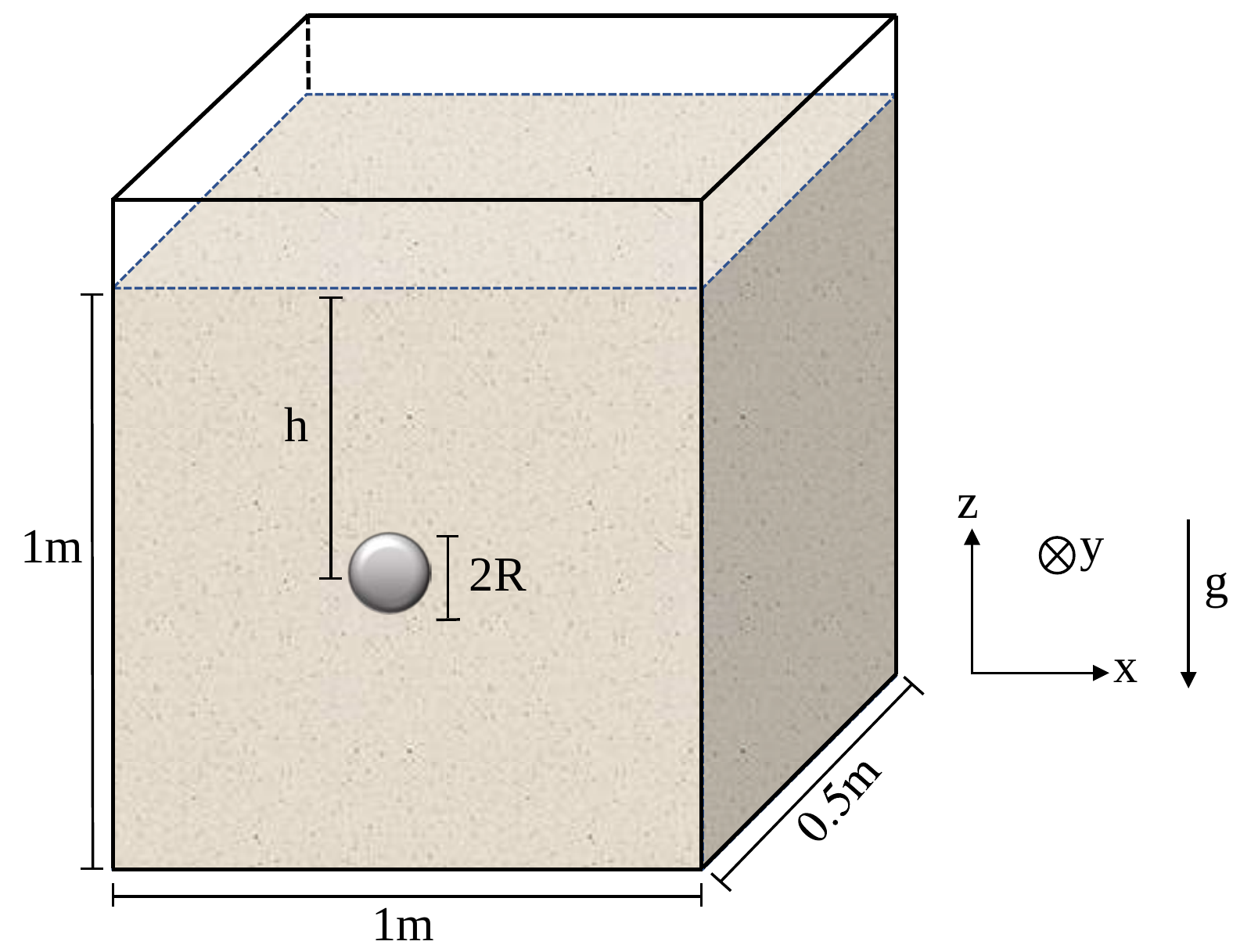}
\caption{The simulated system. Initially, the AO is positioned at $h=h_0$.}
\label{SetUp}
\end{figure}

The dynamics of the AO in the 3D system, sketched in Fig. \ref{SetUp}, were simulated with the open-source software LIGGGHTS \cite{LIGGGHTS}. The AO, modelled as a frictional spherical of radius $R$ and mass density $\rho_m$, was immersed in a granular bed of spheres of diameter $R_0=0.01$m and mass density $2500$Kg/cm$^3$, confined to a container of dimensions $x\times y\times z=1\times0.5\times1.2$m$^3$.
The granular bed filled the space (in metres) $-0.5 \leq x \leq 0.5$, $-0.25 \leq y \leq 0.25$, and $-1.0 \leq z \leq 0$, with the AO's initial depth at $h_0$, measured from the bed's free surface.
We imposed periodic boundary conditions in the $x$ and $y$ directions, a solid bottom at $z=-1.0$m, a free surface at $z=0$m, and Earth gravity $g$ in the $-z$ direction.
The particles interaction potentials were Hertzian and included friction and dissipation. All the simulation parameters are detailed Table 1 in the supplementary material (SM).

The initial state was prepared by first dropping bed particles into the container under gravity, filling it up to $z=-(h_0+R)$, placing the AO as close as possible to $x=y=0$m on the formed surface, and then pouring in the rest of the particles. The total number of particles was $82,340$, yielding a bed particles number density $164,680$m$^{-3}$ and a packing fraction $p=4n\pi R_0^3/3=0.69\pm 0.01$. The effective bed mass density was then $\rho_b=2500\times 0.69=1725$Kg/cm$^3$.
The filling simulation ran until the kinetic energy dissipated to $5\times 10^{-7}$ of the original value.

From this, practically static, state, the AO executed self-energised horizontal oscillations in the $x$-direction, $x(t)=A\sin(\omega t)$, with $0.05$m$\leq A\leq 0.25$m and $1$Hz$\leq f=\omega/2\pi\leq 20$Hz. These ranges were chosen to reduce to a minimum two finite size effects: an oscillatory response of the entire bed to very energetic oscillations and a medium-mediated interaction of the AO with its image in the periodic domain.
The particles trajectories were evolved using the Verlet algorithm with a constant time step of $10^{-5}$s.
The bulk of the simulations were run with an AO radius of $R=0.06$m and density $\rho_m=5000$Kg/m$^3$.
To test effects of several relevant quantities on the dynamics, we ran a large number of simulations, modifying for many combinations of $A$ and $\omega$: (i) AO densities $\rho_m = 1000, 2500, 5000, 7500$Kg/m$^3$; (ii) AO radii $R=0.04,0.06,0.08,0.10$m; (iii) interparticle friction coefficients $\mu = 0.1,0.3,0.5,0.7,0.9$; and (iv) initial depths: $h_0=0.3,0.5,0.7$m.

\begin{figure}[H]
\includegraphics[width=0.52\textwidth]{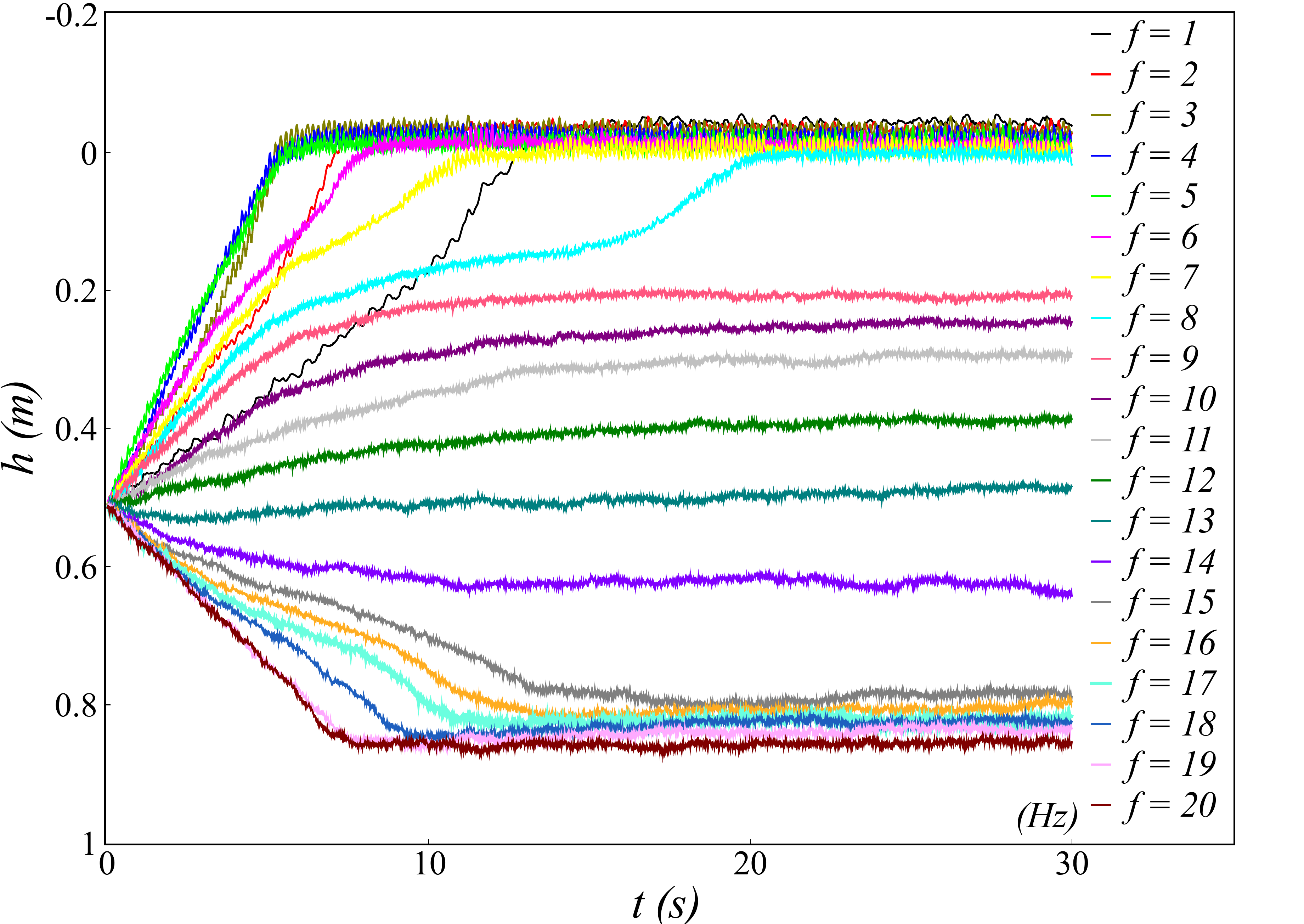}
\caption{The sinking and rising of the AO, at oscillation amplitude $A=0.1$m and frequencies $1$Hz$\leq f\leq20$Hz. 
When $f<9$Hz and $f>14$Hz, the AO reaches the surface and the bottom, respectively. At frequencies $9$Hz$\leq f\leq14$Hz, the AO settles into an equilibrium depth.}
\label{A01Trajectories}
\end{figure}

\begin{figure}[H]
\includegraphics[width=0.52\textwidth]{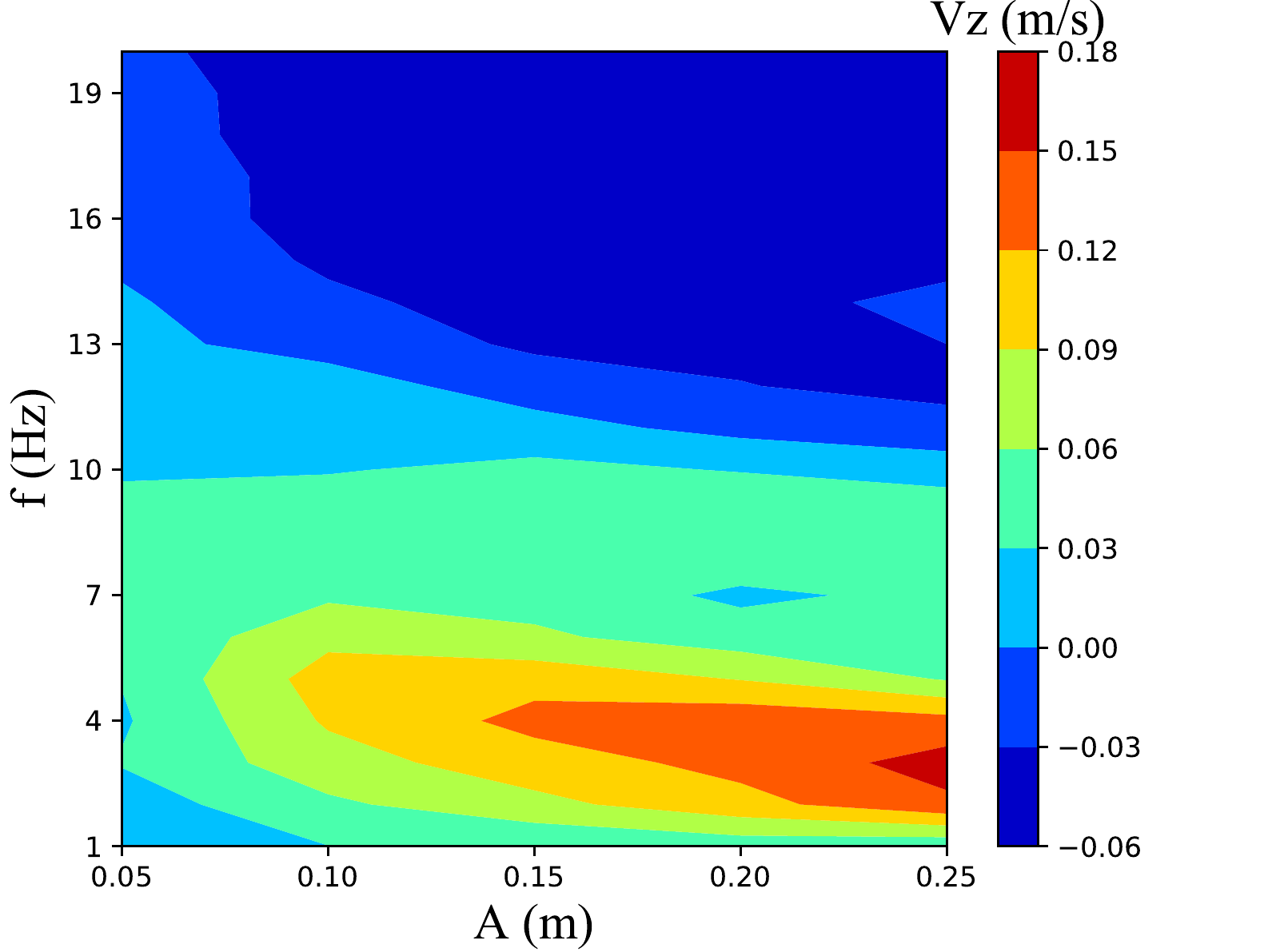}
\caption{ The initial vertical velocity of an AO, starting from $h=0.5$m, plotted in the amplitude-frequency phase space. At low frequencies, the AO rises, with the rising rate increasing until $A$ becomes comparable to the AO's diameter. At high frequencies the AO invariably sinks. }
\label{PhaseDiagram}
\end{figure}

\nin \textbf{Results}:
We observe that the AO may rise, sink or stay at the same depth, depending on $A$ and $\omega$. This confirms that this phenomenon, observed initially in 2D simulations \cite{Huetal16}, extends to 3D.
In Fig. \ref{A01Trajectories}, we plot the AO's depth, $h(t)$, for $A=0.1$m and all frequencies. The phase diagram in Fig. \ref{PhaseDiagram} shows the initial velocity in the amplitude-frequency phase space.


\begin{figure}[H]
\begin{minipage}{0.48\linewidth}
  \centerline{\includegraphics[width=4.6cm]{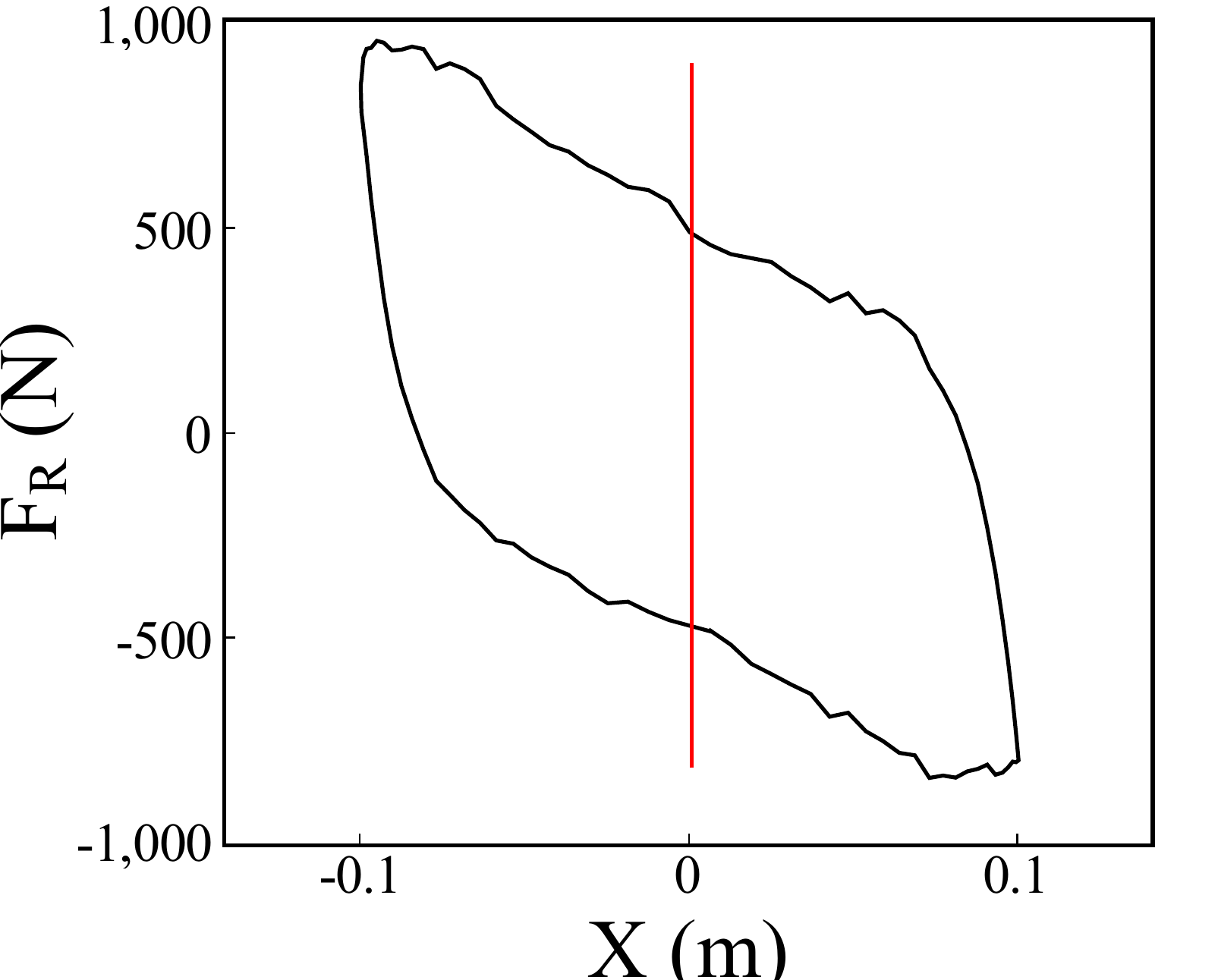}}
  \centerline{  (a)}
\end{minipage}
\hfill
\begin{minipage}{.48\linewidth}
  \centerline{\includegraphics[width=4.6cm]{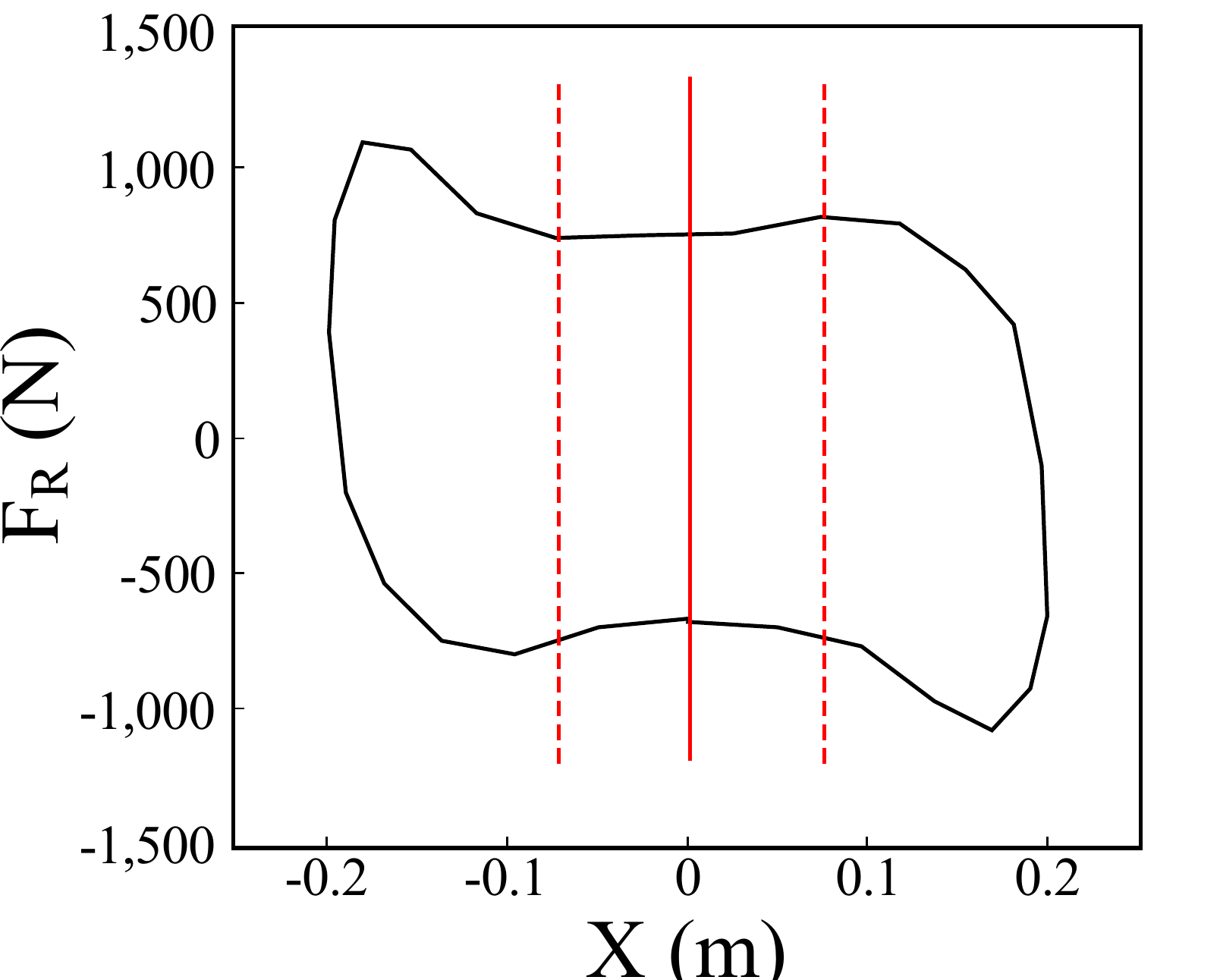}}
  \centerline{  (b)}
\end{minipage}
\caption{Typical examples of the resistance force acting on the AO in the $x$-direction.
(a) The speed never exceeds $v_c$ and the resistance increases linearly during the entire stroke due to the formation of a stagnant zone ahead of the AO.
(b) The speed exceeds $v_c$ in the region marked by the two dashed red lines, leading to the disintegration of the stagnant zone and the resistance force stays constant.}
\label{LinearResistance}
\end{figure}

The simulations lead to the following key observations. \\
\textbf{1.} Although unconstrained in 3D, the AO moves in the $x-z$ plane, with only small fluctuations in the $y$-direction (Fig. 1 in the SM). \\
\textbf{2.} Increasing the frequency generically reduces the rising rate (see Fig. \ref{PhaseDiagram}).\\
\textbf{3.} At low frequencies and all  $A>0.05$m, the AO experiences a linearly increasing resistance force with $x$ throughout a stroke (Fig. \ref{LinearResistance}a) up to a maximum that is proportional to the depth $h$ and independent of $A$ and $\omega$: $F_{R,max}=Ch$, with $C=1770\pm10$N/m (Fig. 2 in the SM).
When $A\omega$ increases above some value, $v_c$, the resistance force stays roughly constant around mid-stroke (Fig. \ref{LinearResistance}b).  \\
\textbf{4.} When $A\omega$ exceeds $v_c$, the AO sinks around the middle of the stroke, with the sinking duration increasing with $A\omega$. The value of $v_c$ increases from $3.1\pm0.1$m/s at depth $h=0.3$m and appears to plateau at $4.2\pm0.5$m/s at $h=0.7$m. \\
\textbf{5.} The rise rate increases with intergranular friction from $\mu=0.1$ and saturates around $\mu\geq 0.5$. \\
\textbf{6.} The sinking rate increases with both the AO's specific density and diameter. \\
Typical plots of these effects are shown in the SM (Figs. 4-6).
In the following, we use these observations to identify the mechanisms driving the vertical motion, model them from first principles, and derive a theoretical EoM for the time-dependent depth, $h(t)$.

\nin \textbf{Theoretical interpretations and modelling}:
The quasi-planar motion of the AO in the $x-z$ plane (observation 1), which is expected from the symmetry of the system, simplifies the analysis and modelling, since the displacements in the $y$-direction take place over much longer time than the oscillation period.
The increase in $F_R$ is intriguing and cannot be explained by simple horizontal drag, which is known to be constant below a critical velocity \cite{Aletal99,Kangetal18}. Rather, it suggests a build-up of a shear-jammed SZ ahead of the AO as it advances. A similar phenomenon is observed when objects penetrate granular media \cite{Bretal13,Kangetal18,Fengetal19}.
The above observations then suggest the following picture. The AO pushes particles forward and out of its way, while shearing the material normal to the motion direction. At low speeds, a jammed SZ cone-like forms, effecting rising via a lift force \cite{Maetal11}, which we calculate below. This force is a result of the local differential hydrostatic-like depth-dependence of the resistance force against the inclined surfaces of the SZ.

It is also known that an SZ build-up occurs only for speeds below a critical velocity \cite{Aletal99}. Thus, at AO higher speeds, the SZ disintegrates, reducing the lift force, which is consistent with observation 4.
Moreover, above the critical velocity not only the SZ but also the entire medium around the AO fluidises, which effects sinking into its supporting layer. This means that the critical velocity and $v_c(h)$ of observation 4 are one and the same.
Fluidisation occurs when: (i) the AO's velocity exceeds $v_c(h)$; (ii) its kinetic energy exceeds the work required to push against the resistance force.
The sinking rate is expected to increase linearly with the excess energy kinetic energy and, therefore, the sinking and rising rates depend differently on $A$ and $\omega$. This leads to an intricate competition, modulated by AO's speed and excess energy in different parts of the stroke.
A rise-sink competition has been also noted and discussed in 2D simulations \cite{Huetal16}, but its driving mechanisms, described above, were not identified there.
In particular, our observations undermine the cavity model \cite{Huetal16}, which presumes that the AO climbs over particles falling into the cavity left in its wake. The cavity model predicts a uniform rise throughout the stroke, while we observe rise at low velocities and sink at high ones, within the same stroke.

The above picture is the basis for the following model. When $A\omega\leq v_c$, $v_{sink}=0$.
When $A\omega>v_c$, the AO may only sink when $-\tau\leq t\leq\tau$, with $\cos{\omega\tau}=v_c/A\omega$. Whether it sinks during this interval or not depends on
the available excess energy for fluidisation, $\delta E = \beta mv^2(t)/2 - W$, where $W$ is the work done against the resistance force and $\beta<1$ is the fraction of the kinetic energy invested in fluidising the support layer. Assuming equal fluidisation of the medium in $y$ and $z$ directions, we set $\beta=1/4$.
The work depends on the resistance force, which can be modelled using observations 3 and 4 and $x=A\sin{\omega t}$:
\beq
F_R = \frac{Ch}{2}
\begin{cases}
1\! +\!  \sin{\omega t} & A\omega\!\leq\!v_c \\
1\!  +\!  \sin{\omega t} & A\omega\!>\!v_c\ ;\  t\leq-\tau \\
1\!  -\!  \sin{\omega\tau} & A\omega\!>\!v_c\ ;\  -\tau\! <\!  t\!  <\!  \tau \\
1\!  -\!  2\sin{\omega\tau}\! +\! \sin{\omega t} & A\omega\!>\!v_c\ ;\  \tau\! \leq\!  t\ .  \\
\end{cases}
\label{ResForce}
\eeq
The AO can sink only between $-\tau < t < \tau$ when $\delta E > 0$. To find the duration of the sink phase, we solve for $t_u$ when $\delta E(t_u) = 0$. We find
\beq
\omega t_u = \arcsin\frac{\sqrt{1+4\gamma(\gamma - \sin{\omega\tau})}-1}{2\gamma} \ ,
\label{tu}
\eeq
where $\gamma\equiv\beta mA\omega^2/\left[Ch(1 - \sin{\omega\tau})\right]$ is a dimensionless measure of the AO's kinetic energy.
At low energies, when $\gamma<\sin{\omega\tau}$, $-T/4<t_u<-\tau$ there is no sinking and the AO rises only outside the interval $-\tau\leq t\leq\tau$.
When $\gamma\geq\sin{\omega\tau}$ and $t_u\leq\tau$, the AO sinks within the interval $-\tau\leq t\leq t_u$, and neither sinks nor rises during $t_u\leq t\leq \tau$.
When $\gamma\geq\sin{\omega\tau}$ and $t_u>\tau$, the AO sinks within the entire interval $-\tau\leq t\leq\tau$.
Thus, there are four possible dynamics, one when $A\omega\leq v_c$, which call case A, and the above three when $A\omega> v_c$, which depend on the relation between $t_u$ and $\tau$ and which we call BI, BII and BIII, respectively.

The sinking depth per stroke is expected to be proportional to the excess energy and inversely proportional to the depth because of the increased difficulty in fluidisation with hydrostatic-like pressure, $\delta h_{sink}=C_{sink}\delta E/h$. Our simulations suggest that $C_{sink}\approx 1\times10^{-4}$s$^2$/Kg. Integrating $\delta E/(t_u+\tau)$ between $-\tau$ and $t_u$ we find the total excess energy in this interval and the sinking per stroke in this regime is
\beq
\begin{split}
&\Delta h_{sink} = \frac{C_{sink}CA(1 - \sin{\omega\tau})}{2}\times \\
&\times\left[\frac{\gamma}{2} - \sin{\omega\tau}  + \frac{\sin{2\omega \theta} + \sin{2\omega\tau} }{4\omega(\theta + \tau)}
- \frac{\cos{\omega \theta} - \cos{\omega\tau}}{\omega(\theta + \tau)} \right]
\end{split}
\label{Hsink}
\eeq
with $\theta=t_u$($\tau$) when $t_u<\tau$ ($t_u\geq\tau$).

Rising is caused by conversion of the resistance into an upward lift by the inclined surface of the SZ. The SZ is generically conical \cite{Kangetal18,Fengetal19,Bretal13} and, for simplicity, we model the AO and the SZ as joined cone and a half-sphere (Fig. 3. in the SM). The resistance force is proportional to the hydrostatic-like pressure, $K\rho_bgz$ \cite{Kangetal18,Fengetal19}, and the total resistance, calculated in detail in the SM, is
\beq
F_R = \pi K\rho_bghR^2\sin{\alpha} \ .
\label{ResForce1}
\eeq
We also detail in the SM a first-principles calculation of the lift force:
\beq
F_z = \frac{\pi R^3K\rho_bg\cos{\alpha}\sin^2{\alpha}}{3} \ .
\label{LiftForce}
\eeq
From (\ref{ResForce1}) and (\ref{LiftForce}), we have
\beq
\mid\! F_z\!\mid = \frac{\sin{2\alpha}}{6}\left(\frac{R}{h}\right)\mid\! F_R\!\mid \ .
\label{LiftForce1}
\eeq
By construction, $\alpha\leq\pi/4$ and too sharp cones blunt against the resistance. Previous works show $\pi/6 < \alpha$ \cite{Vietal15,Fengetal19,Haetal20}. Thus, $\sqrt{3}/2\leq\sin{2\alpha}\leq1$ and, for our purpose, can be approximated as a constant, $0.93\pm0.07$.

\begin{figure}[H]
\includegraphics[width=0.5\textwidth]{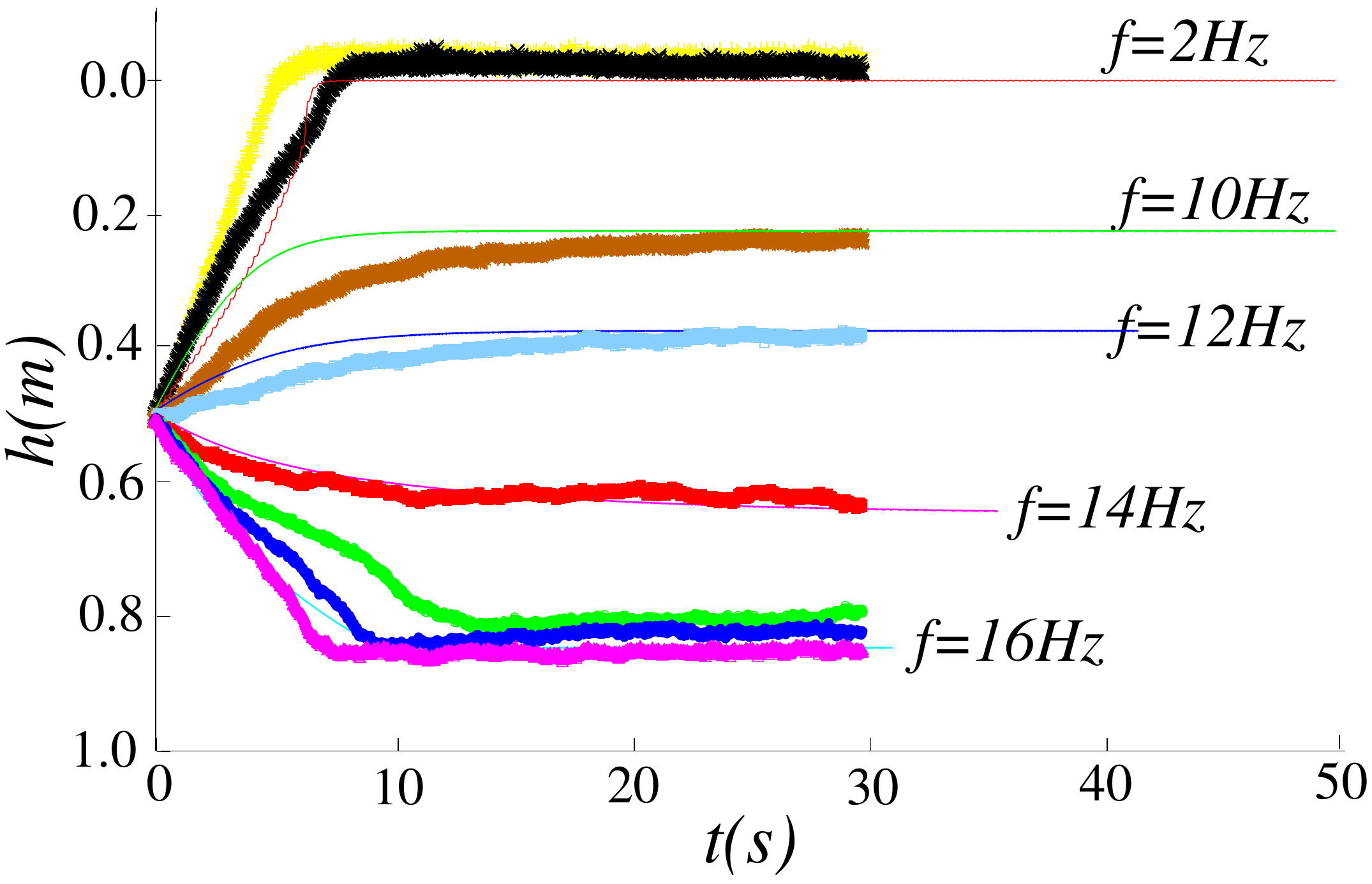}
\caption{A comparison of the theoretical predictions and the numerical simulations for $A=0.1$m, as a typical example. Shown are all the frequencies for which equilibrates within the system, as well as choice frequencies at which the AO reaches either the surface or the bottom. The theory captures well the main trend at all frequencies, the equilibration depth, and the rising/sinking initial speeds. The latter is somewhat overestimated at frequencies $10$Hz and $12$Hz, which we attribute to the system finite size. }
\label{NumericalSolutionsCompare}
\end{figure}

$F_z$ lifts not only the AO but also a part $\eta$ $(<1)$ of the column of particles above it because of sideways flow. Altogether, the lifted mass is
\beq
M = \frac{2\pi R^3\rho_b}{3}\left[\left(\frac{3h}{2R} - 1\right)\eta + 2\chi \right] \ ,
\label{LiftMass}
\eeq
where $\chi\equiv\rho_m/\rho_b$. Aiming to model the dynamics away from the surface, we set $\eta=1$.
Using (\ref{ResForce}), (\ref{LiftForce1}) and (\ref{LiftMass}) to integrate $F_z/M$ twice gives the rise. When $A\omega\leq v_c$ the rise during the forward stroke is
\beq
\Delta h_{rise} = \frac{CR\sin{2\alpha}}{24M\omega^2}\left(\pi^2 - 4\right) \ .
\label{RiseHA}
\eeq
When $A\omega>v_c$, the AO rises during the forward stroke only outside the region $-\tau\leq t\leq\tau$:
\beq
\Delta h_{rise} = \frac{CR\sin{2\alpha}}{24M\omega^2} \left[\lambda \left(\frac{\pi}{2} - \omega\tau\right)^2 - 2(1 - \sin{\omega\tau})\right] \ ,
\label{RiseHBC}
\eeq
with $\lambda=1$ for the rise between $-T/4$ and $-\tau$ and $\lambda=1-\sin{\omega\tau}$ between $\tau$ and $T/4$.
Using (\ref{Hsink}), (\ref{RiseHA}), and (\ref{RiseHBC}), gives the depth EoM, $dh/dt$:
\beq
\frac{dh}{dt} = \frac{\omega}{\pi}\left(\Delta h_{sink} - \Delta h_{rise} \right) \ .
\label{EoM}
\eeq
The nonlinear dependencies of $\Delta h_{sink}$ and $\Delta h_{rise}$ on $A$, $\omega$ and $h$, make a general analytic solution difficult.
A careful inspection of the EoM reveals that, when both terms are finite, the AO settles eventually at an equilibrium depth, unless the AO reaches the surface or the bottom. This conclusion is supported by the numerical solutions below.
In Fig. \ref{NumericalSolutionsCompare}, we compare the simulations and the solutions for $h(t)$ for $A=0.1$m and the frequencies, $2$Hz$\leq f\leq 16$Hz. 
The theory captures well which frequencies equilibrate within the system and the equilibration depths. It also models well almost all the initial rising and sinking rtes, except for some overestimate of the rise velocity for $10$Hz and $12$Hz, which we believe is caused by the system finite size.

\nin \textbf{Discussion and conclusions}:
We have considered an important case study of the interaction of active objects (AOs) with passive matter - the sinking and rising of an active self-energised object, horizontally oscillating in a granular medium. Such dynamics are relevant to modelling lizards burrowing in sand, robotic locomotion in granular environments, and survival strategy in quicksand.
We have identified the two competing mechanisms that give rise to intricate vertical dynamics: rising at low speeds, enabled by the formation of a jammed stagnant zone (SZ) ahead of the AO, and sinking when the kinetic energy is sufficiently high to fluidise the layer supporting it.
Based on these mechanisms, we have developed a first-principles theory for the depth equation of motion.

The SZ develops ahead of the AO only below a critical velocity, which is slightly depth-dependent. Its cone-like structure converts the force resisting the motion into a lift force. We calculate the lift force in different regimes of the amplitude-frequency phase space and from it the rate of rising.
The sinking rate, which is proportional to the excess energy available to fluidise the AO's supporting layer, has also been calculated in the different regimes. Combining the rising and sinking rates, we constructed the dependence of the depth equation motion on the oscillation parameters.

The theory is supported with an extensive range of simulations, in which we vary oscillation amplitudes and frequencies, initial depth, inter-particle friction coefficient, and AO properties. A phase diagram has been constructed for the rise-sink behaviour in the amplitude-frequency plane and the theoretically predicted trajectories agree well with the numerical simulation results, as shown in Fig. \ref{NumericalSolutionsCompare}. The theory overestimates slightly the initial rising rate at the two frequencies for which the AO equilibrates inside the system. We believe that this is because the theory applies strictly away from the surface, which requires larger simulations than we could run. 

Nevertheless, although our theory applies well away from the surface, it is interesting to check its predictions against real lizards. The Uma Scoparia, whose head mass is about $4$g, burrows in sand by oscillating its head at $A\approx0.02$m and $f\approx30$Hz $\approx190$s$^{-1}$ \cite{Lizard}.  At the surface, $v_c\to 0$ and almost all the vibration kinetic energy goes to fluidise the supporting layer. Using our deep-medium estimate of $C_{sink}$ our solution suggests that it burrows into the and at a rate of $4$cm/s. With a body height of about $2$cm, this means that its head can disappear in less than a second, which agrees with observations \cite{Lizard}.

\begin{acknowledgments}
This work was supported by the Fundamental Research Funds of NUDT, Grant No. ZK16-03-01.
\end{acknowledgments}


\end{document}